# Vacancy complexes in nonequilibrium germanium-tin semiconductors

S. Assali,[1,*] M. Elsayed,[2,3] J. Nicolas,[1] M. O. Liedke,[4] A. Wagner,[4] M. Butterling,[4] R. Krause-Rehberg,[2] and O. Moutanabbir[1,*]

[1]Department of Engineering Physics, École Polytechnique de Montréal, C. P. 6079, Succ. Centre-Ville, Montréal, Québec H3C 3A7, Canada
[2] Martin Luther University Halle, 06099 Halle, Germany
[3]Department of Physics, Faculty of Science, Minia University, 61519 Minia, Egypt
[4] Helmholtz-Zentrum Dresden-Rossendorf, Institute of Radiation Physics, Bautzner Landstraße 400, 01328 Dresden, Germany

**ABSTRACT:**

Understanding the nature and behavior of vacancy-like defects in epitaxial germanium-tin (GeSn) metastable alloys is crucial to elucidate the structural and optoelectronic properties of these emerging semiconductors. The formation of vacancies and their complexes is expected to be promoted by the relatively low substrate temperature required for the epitaxial growth of GeSn layers with Sn contents significantly above the equilibrium solubility of 1 at.%. These defects can impact both the microstructure and charge carrier lifetime. Herein, to identify the vacancy-related complexes and probe their evolution as a function of Sn content, depth-profiled pulsed low-energy positron annihilation lifetime spectroscopy and Doppler broadening spectroscopy were combined to investigate GeSn epitaxial layers with Sn content in the 6.5-13.0 at.% range. The investigated samples were grown by chemical vapor deposition method at temperatures between 300 and 330 °C. Regardless of the Sn content, all GeSn samples showed the same depth-dependent increase in the positron annihilation line broadening parameters, relative to that of epitaxial Ge reference layers. These observations confirmed the presence of open volume defects in as-grown layers. The measured average positron lifetimes were found to be the highest (380-395 ps) in the region near the surface and monotonically decrease across the analyzed thickness, but remain above 350 ps. All GeSn layers exhibit lifetimes that are 85 to 110 ps higher than those recorded for Ge reference layers. Surprisingly, these lifetimes were found to decrease as Sn content increases in GeSn layers.

These measurements indicate that divacancies are the dominant defect in the as-grown GeSn layers. However, their corresponding lifetime was found to be shorter than in epitaxial Ge thus suggesting that the presence of Sn may alter the structure of divacancies. Additionally, GeSn layers were found to also contain a small fraction of vacancy clusters, which become less important as Sn concentration increases. The interaction and possible pairing between Sn and vacancies have been proposed to explain the reduced formation of larger vacancy clusters in GeSn with higher Sn content.

# I. INTRODUCTION

Developing chip-scale lasers, detectors and other active optoelectronic devices on silicon provides a viable path for monolithic manufacturing of photonic integrated circuits with significant cost reduction and new functionalities for a wide range of applications.[1] In this regard, Sn-containing group IV semiconductors (Si)GeSn have been suggested as potential building blocks for silicon-compatible light sources.[2] This material system provides two degrees of freedom for band structure engineering, namely, alloying and strain. This characteristic is central to engineer novel low-dimensional systems and heterostructures in a similar fashion to the mature III-V semiconductors. Moreover, unlike the indirect band gap Si and Ge, a direct fundamental band gap can be obtained in (Si)GeSn which is promising for efficient emission and detection of light.[3–5] The flexibility in band gap engineering provided by these group IV semiconductors has sparked a recent surge of interest in the growth of device quality layers on Si or Ge substrates.[6–9] The nature of band gap and its energy depend on the composition and strain in the alloy.[3–5]

The epitaxial growth of (Si)GeSn semiconductors has been proven to be very challenging due to the low solubility of Sn in Si (<0.1 at.%) and Ge (<1 at.%). For instance, in the $Ge_{1-y}Sn_y$ binary system only one component alloy is allowed in equilibrium near each element: $Ge_{1-y}Sn_y$ with y < 1 at.% or y > 99.4 at.%.[10,11] Additionally, only the diamond structure of α-Sn is semiconducting, which is stable below 13.2 °C. Above this temperature, α-Sn transforms to the metallic phase β-Sn, thus adding to the long list of the growth hurdles. The recent progress in controlling the growth kinetics overcame these difficulties leading to the epitaxial growth of device-quality binary and ternary layers with high Sn content.[6–9,12] However, to avoid surface segregation and phase separation, the growth is typically carried out below 400°C, thus raising fundamental questions regarding point defects that are inherent to low growth temperatures. In fact, the presence of native point defects such as vacancies and vacancy complexes can greatly impact the microstructure and the electronic properties of the epitaxial layers. For instance, vacancies can promote the motion and interaction of dislocations thus impacting both the density and the type of extended defects.[13] Moreover, vacancy-related lattice defects can also strongly limit the efficiency of light emission and detection via doping compensation and carrier trapping.[14–16] Despite their importance, studies of the vacancy-related complexes in epitaxial Sn-containing group IV semiconductors are still conspicuously missing in literature. With this perspective, this work reports on direct analysis of vacancy complexes and elucidates their behavior as a function of Sn content in metastable GeSn epitaxial layers. By combining depth-profiled positron annihilation lifetime spectroscopy (PAS) and Doppler broadening measurements, we found that divacancy is the dominant complex observed in GeSn layers independently of Sn content in the investigated range of 6.5-13 at.%. The as-grown layers were also found to contain a small fraction of larger vacancy clusters. Intriguingly, the increase in Sn incorporation was found to be associated with a slight increase in divacancies and concomitant decrease in vacancy clusters.

## II. EXPERIMENTAL DETAILS

The investigated samples were grown on a 4-inch Si (100) wafers in a low-pressure chemical vapor deposition (CVD) reactor using ultra-pure $H_2$ carrier gas, and 10 % monogermane ($GeH_4$) and tin-tetrachloride ($SnCl_4$) precursors, following a recently developed growth protocol.[8,12,17] First, a 500-700 nm-thick Ge-VS was grown at 450 °C, followed by thermal cyclic annealing (>800 °C) and additional Ge deposition. Next, a 500-700 nm-thick GeSn layer (Fig. 1a) was grown at a fixed temperature in the 300-330 °C range, corresponding to a Sn content in the 6.5-13 at.% range, as estimated from Reciprocal Space Mapping (RSM) X-ray diffraction (XRD) (Fig. 1a, inset). Herein, the highest Sn content is obtained at the lowest growth temperature and *vice versa*. Note that a ~60 nm-thick GeSn buffer layer grown at 320 °C was used for the subsequent growth of the 310 °C and 300 °C samples. The residual (compressive) in-plane strain in the GeSn layers being lower than -0.3 % allows for a direct comparison in the 2θ-ω scans around the (004) XRD order in Fig. 1b, where a shift of the GeSn peak toward lower angles with increasing Sn content is observed.[12] PAS analysis was carried out to probe open volume defects in different as-grown GeSn layers. By performing PAS lifetime measurements, the change in the annihilation parameters is measured when positrons are trapped at vacancy-like defects and the nature of these open volume defect is obtained from lifetime measurements. The Doppler broadening of the positron annihilation line allows the identification of vacancies and their complexes by exploiting the high sensitivity to the chemical surrounding of the vacancy-like defects.[18] The combination of lifetime and Doppler broadening measurements then yields a qualitative assessment of vacancy complexes and their size.

Doppler broadening analyses were carried out using a variable energy positron beam.[19,20] Monoenergetic positrons were produced by a 20 mCi $^{22}$Na source assembled in transmission with a 1 μm monocrystalline tungsten moderator and transported in a magnetic guidance system to the sample. The beam is characterized with a diameter of 4 mm and an intensity of $5\times10^2$ e$^+$/s. The samples were measured at room temperature under ultra-high vacuum. A high-purity Ge detector with an energy resolution of (1.09±0.01) keV at 511 keV was used to record the annihilation spectra. A spectrum of about $5\times10^5$ counts in the 511 keV peak is acquired at each positron kinetic energy $E$. The Doppler broadening spectrum of the 511 keV annihilation line is characterized by two parameters, $S$ (low electron momentum fraction) and $W$ (high electron momentum fraction).

The trapping of positrons at open-volume defects is detected as an increase in S parameter and a decrease in W parameter (not shown here). Using positron energies in the range of 0.03–35 keV the samples are analyzed up to a mean penetration depth of ~2 μm below the surface to ensure that the entire thickness of GeSn layers (700 nm) is probed. Positron annihilation lifetime spectroscopy (PALS) measurements were performed using the mono-energetic positron source (MePS) at the superconducting electron linear accelerator ELBE (electron LINAC with high brilliance and low emittance) located at the Helmholtz-Zentrum Dresden-Rossendorf. The working principles of this facility are detailed in Ref. 21. The lifetime measurements were done at room temperature. $10^7$ counts were accumulated in each positron lifetime spectrum. The lifetime spectra were decomposed in two components, $n(t) = (I_1/\tau_1)e^{-t/\tau_1} + (I_2/\tau_2)e^{-t/\tau_2}$, convoluted with the Gaussian resolution function of the spectrometer using the lifetime program LT9.[22] The average lifetime is determined from the lifetime deconvolution, $\tau_{av} = I_1\tau_1 + I_2\tau_2$. Here, $\tau_i$ and $I_i$ are the positron lifetime and its relative intensity, respectively, of each lifetime component i. An average lifetime above the bulk value $\tau_b$ is a sign of the existence of vacancy defects in the material. This parameter can be experimentally observed with high accuracy and a change as small as 1 ps in its value can be reliably measured.[23,24]

## III. RESULTS

The evolution of the S-parameter as a function of the positron energy in each GeSn layer is shown in Figure 2. For comparison, a reference measurement performed on a 600 nm-thick Ge-VS is also shown. We note that the penetration depth of the positron beam was estimated at different energies using Ge lattice as a reference. It is noticeable that the S-parameter increases as the positron energy increases up to 10 keV ($z_{mean}$~310 nm). At higher depths (E>12 keV), the positron beam reaches the Si wafer and a constant value of $S$~0.52 is recorded. All GeSn samples show a similar trend. The $S$-parameter is larger than in the case of the Ge-VS up to 10 keV ($z_{mean}$~310 nm), followed by a progressive decrease reaching the Si value at 30 keV ($z_{mean}$~2000 nm). We note that despite the use of higher energies, the interaction of the positron beam with the GeSn layers induced by the broadening of the Makhov profile,[25] leads to $S$-parameter values in the underlying Ge-VS larger than in the reference layer without GeSn. Fig. 1b displays the $S$-parameter for GeSn layers normalized to that of Ge-VS ($S/S_{Ge-VS}$). In principle, the increase in the normalized $S/S_{Ge-VS}$ parameter results from the reduced diffusion length of the positrons due to an increased density of defects.[26] Thus, a higher amount of open-volume defects is observed when moving from the surface of the GeSn layer down to the interface with the Ge-VS. Typically, when the defect concentrations does not vary significantly, the values of the normalized $S$-parameter are expected to be 1.030, 1.034–1.044, 1.05–1.07, and 1.10–1.14 for monovacancies, divacancies, small vacancy clusters, and voids, respectively.[18,27,28] Note, however, that the normalization is done here using Ge-VS as reference and not a bulk material, which should underestimate $S/S_{Ge-VS}$ values as Ge-Vs contains more open volume defects as compared to bulk Ge. Nevertheless, it is reasonable to infer that the data displayed in Fig. 2b provide the first hints that the predominant defects in as-grown GeSn layers are likely divacancies and vacancy clusters, as confirmed below. At energies above 12 keV ($z_{mean}$>600 nm), the beam reaches the underlying Ge-VS and a clear decrease in $S/S_{Ge-VS}$ is observed, eventually evolving toward unity when approaching the Si substrate.

The positron lifetime measurements as a function of depth are exhibited in Fig. 3a. In the Ge-VS, the average lifetime $\tau_{AV}$ drops from 360 ps near the surface down to 315 ps at the interface with Si substrate. Independently of the probed depth, the average lifetime values are well above the bulk Ge lifetime $\tau_b$=224-228 ps,[29,30] thus confirming that vacancy-related defects are present in the Ge-VS. For GeSn layers, a decrease in $\tau_{AV}$ with increasing depth is observed in all samples. It is noticeable that the incorporation of Sn is always associated with an increase in the average lifetime as compared to Ge-VS, but with a less pronounced decrease as a function of depth. The effect of composition is summarized in Fig. 3b, showing the average lifetime $\tau_{AV}$ at various depths as a function of Sn content. Regardless of the mean depth considered, $\tau_{AV}$ always increases in presence of Sn. Surprisingly, the sample with the highest Sn content (13 at.%) always shows the shortest average lifetime across the whole GeSn thickness. At a first glimpse, this observation seems rather counterintuitive as the layers with the highest Sn content were grown at a slightly lower temperature (300 °C), which should further promote the formation of vacancies. As it will be discussed later, a possible interaction between Sn and vacancy complexes can be invoked to explain this behavior.

To investigate the nature of the detected open volume defects, the lifetime spectra were fitted with two lifetime components.[18] For the reference Ge-VS, the results for the lifetimes $\tau_1$ and $\tau_2$ and the intensity $I_2$ associated with the second component are plotted as a function of the mean depth in Fig. 4. The intensity for the longer lifetime component $I_2$ is always higher (55-75 %) than the short-lived component $I_1$ across the whole layer thickness. Interestingly, the $\tau_1$ values are in 250-225 ps range, hence at much shorter-lived values compared to the 405-350 ps range observed for $\tau_2$. The calculated lifetime values in bulk Ge for the isolated vacancy is 263 ps, while for divacancies the lifetime increases to 325 ps.[29] Experimentally, the reference bulk Ge lifetime is $\tau_b$=224-228 ps.[29,30], thus the $\tau_1$ values are approaching the defect-free bulk lifetime. In fact, a $\tau_2$>330 ps was recorded for neutral divacancy in bulk Ge samples irradiated with neutrons,[31] while values below 320 ps are generally associated with negatively charged divacancies, which results in a background p-type doping in Ge.[15] In proton-irradiated Ge bulk samples, vacancy-like defects with a negative charge state and a lifetime of 294 ps, assigned as V–As complexes, were also reported.[32] However, it is noteworthy that monovacancies are usually measurable only at temperatures lower than 200K, since above this temperature neutral monovacancies paired to form divacancies.[33] In addition, clusters formed by 3 or more vacancies are believed to be associated with a lifetime longer than 365 ps.[34] The deconvoluted lifetime results in Fig. 4 shows that clusters made of multiple vacancies[33] are present with a $\tau_2$=405-350 ps. However, the $\tau_1$=250-225 ps is much lower than the expected values for divacancies, thus the higher growth temperature of 450 ºC in combination with the thermal cyclic annealing above 800 °C seems to contribute to the annihilation of most of the divacancies, while the remaining tend to form vacancy clusters. This is consistent with an early report on divacancy clustering in neutron-irradiated Ge.[29]

For GeSn samples, the results for the lifetimes $\tau_1$ and $\tau_2$ and their corresponding intensities $I_1$ and $I_2$ as a function of the mean depth at Sn content of 13 at.% are plotted in Fig. 5a-b. Note that in the samples with lower Sn contents the analysis yields qualitatively the same behavior. The shortest lifetime $\tau_1$ decreases from 366 ps near GeSn surface (20 nm) to 324 ps at the interface with the Ge-VS (623 nm). A striking difference is, however, observed for the second lifetime $\tau_2$, where much larger values ranging from 490 ps to 750 ps are obtained. $\tau_2$ remains practically unchanged with a value around 500 ps recorded at a depth between 50 and 200 nm. It is remarkable that, across the whole GeSn layer thickness, the shortest lifetime $\tau_1$ dominates by far the recombination process, with an intensity $I_1$ being always higher than 80 %.

Let us now examine the effect of the Sn content. Representative sets of intensities and lifetimes recorded as a function of Sn composition at a fixed mean depth of 215 nm are plotted in

Fig. 5c-d. It is noticeable that the intensity $I_1$ increases from 65 % to 83 % with increasing Sn content, while the shortest lifetime $\tau_1$ remains constant at 335±2 ps. At the same time, the longer lifetime $\tau_2$ increases slightly from 469 ps to 489 ps with increasing Sn content in the layer. In GeSn the shortest lifetime $\tau_1$=335 ps can be associated with the presence of divacancies in the GeSn layer, while the longest lifetime $\tau_2$ >450 ps is associated with clusters made of multiple vacancies, most likely with more than 5 vacancies.[33] However, the intensity $I_2$ being much lower than $I_1$ demonstrates that the concentration of vacancy-clusters is rather low compared to that of divacancies. Indeed, the increase of intensity $I_1$ with Sn content, reaching 80% at 13 at.% of Sn (Fig. 5c), indicates that divacancies are the predominant type of point defects in GeSn. This result also provides clear evidence that divacancy is stable in GeSn at room temperature, which is the temperature at which the positron annihilation measurements were conducted. We note that the GeSn alloys composition was controlled during growth by varying the temperature of the samples from 330 °C (6.5 at.%) down to 300 °C (13 at.%). Here, one may argue that lowering the growth temperature may limit the mobility of vacancy complexes and consequently their interaction can become less probable, thus reducing the likelihood of the formation of larger vacancy clusters. However, it is important to point out that that the change in growth temperature from 300 to 330 ºC is too small to significantly impact the dynamics of vacancy complexes and their interactions. For instance, the migration of a divacancy in Ge involves activation energies above 1eV regardless of the chosen pathways (direct migration of divacancies or migration after split to form the highly mobile single vacancies).[35] Therefore, it is reasonable to assume that the change in thermal energy under our conditions may have a negligible effect.

As mentioned above, $\tau_1$ does not depend on the Sn composition in the investigated range and is always more than 85 ps higher than the 250-225 ps recorded for the Ge-VS. This may suggest that the divacancy structure is perhaps altered by the presence of Sn and the associated lattice stress. In general, it is reasonable to believe that because Sn atom is larger than Ge atom the pairing of Sn with vacancies would be a viable path to locally minimize the stress associated with Sn incorporation. It is currently well established that Sn occupies substitutional sites in GeSn lattice and is randomly distributed[8] with no sign of any short-range ordering.[36] However, early reports on the interaction of Sn with simple defects in irradiated Ge provided evidence of an affinity of Sn atoms to attract vacancies in order to relief the stress in Ge lattice.[37] This attractive interaction between Sn and vacancies was also suggested based on *ab initio* calculations combined with emission channeling studies of Sn-implanted Ge.[38] Based on thermodynamic considerations, these studies confirmed that the isovalent Sn expectedly favors substitutional sites, but it can also trap vacancies. After trapping vacancies, Sn evolves to occupy the bound centered site under vacancy-split configuration. Note that the formation of Sn-divacancy complexes has also been reported in irradiated Ge, but it was shown to anneal out at 220K,[39] which is significantly below the growth temperatures used in this study. Nevertheless, the formation of these complexes in epitaxial films cannot be ruled out as the growth kinetics can affect their thermal stability.

Our lifetime measurements agree at least qualitatively with the aforementioned observations. Indeed, the results point out to a possible effect of Sn that can locally *freeze* vacancies thus limiting their clustering in the grown layers. Indeed, the observed drop in the intensity of vacancy clusters $I_2$ as Sn content increases underscores a certain role of Sn in reducing vacancy migration and interaction to form clusters. This effect yields a decrease in $\tau_{AV}$ as the long-lived $\tau_2$ component becomes less important at higher Sn incorporation. Finally, it is important to emphasize that the dislocations in the proximity of the GeSn-Ge interface[8,12] do not have an impact on both Doppler broadening and on the lifetime measurements. Dislocations are either free of vacancies or can contain vacancies. Vacancy-free dislocations would act as shallow positron traps only at low temperatures (T<100K), hence irrelevant under our conditions. They show a lifetime close to the bulk value at room temperature. If dislocations contain vacancies they would act as deep positron trap, leading to an increase in the (normalized) *S*-parameter to 1.02 and in a lifetime lower than 300 ps. Here, the experimental lifetime recorded above 320 ps excludes the presence of monovacancies in GeSn when measured at room temperature.

**CONCLUSION**

Vacancy-related complexes were investigated in GeSn semiconductors with Sn content in the 6.5-13.0 at.% range using depth-profiled pulsed low-energy positron annihilation lifetime spectroscopy and Doppler broadening spectroscopy. Regardless of the Sn content, an increase in the broadening parameter, relative to that of epitaxial Ge reference layers, with increased probed depth was observed in all samples. This indicates the presence of open volume defects in the epitaxially-grown layers. A monotonic decrease in the measured lifetime from 380-395 ps (close to the sample surface) down to 350 ps (in the proximity of the Ge-VS) was recorded across the GeSn thickness. When compared to the Ge-VS reference layers, the measured average lifetimes in GeSn were found to be 20 to 140 ps higher. In addition, the average lifetime was also found to be inversely proportional to Sn content in the GeSn layers. These results show that divacancies are the dominant defects in the as-grown GeSn layers, while only a small fraction of vacancy clusters is present in the alloy, with a further reduced amount as the Sn content of the alloy increases. A possible attractive interaction between Sn and vacancies has been suggested to explain the reduced clustering of vacancies in GeSn as Sn content increases.

## ACKNOWLEDGMENTS

The authors thank J. Bouchard for the technical support with the CVD system. O.M. acknowledges support from NSERC Canada (Discovery, SPG, and CRD Grants), Canada Research Chairs, Canada Foundation for Innovation, Mitacs, PRIMA Québec, and Defence Canada (Innovation for Defence Excellence and Security, IDEaS). S.A. acknowledges support from Fonds de recherche du Québec-Nature et technologies (FRQNT, PBEEE scholarship). Positron annihilation experiments were carried out in the ELBE facility thanks to the large infrastructure program of the EU (proposal no. POS18101148). We acknowledge BMBF for the PosiAnalyse (05K2013) grant, the Impulse- und Networking fund of the Helmholtz-Association (FKZ VH-VI-442 Memriox), and the Helmholtz Energy Materials Characterization Platform (03ET7015).

## AUTHOR INFORMATION

Corresponding Authors:

*E-mail: simone.assali@polymtl.ca

*E-mail: oussama.moutanabbir@polymtl.ca

Notes:

The authors declare no competing financial interest.

**Figure 1.** (a) Cross-sectional TEM image along the [110] zone axis. Inset: RSM map showing the graded composition up to 12.9 at.% and the 13.7 at.% uniform growth on top in the 13 at.% GeSn sample. (b) 2θ-ω scans around the (004) X-ray diffraction order acquired on the 6.5-13 at.% GeSn samples (grown between 300-330 °C) and the Ge-VS substrate (grown at 450 °C and annealed >800 °C).

**Figure 2**. (a) S-parameter as a function of the incident positron energy for all GeSn samples and for the reference Ge-VS. (b) Normalized $S/S_{Ge-VS}$ parameter as a function of the incident positron energy for all GeSn samples.

**Figure 3**. (a) Average lifetime $\tau_{AV}$ as a function of the positron mean depth for all GeSn and Ge-VS samples. (b) $\tau_{AV}$ as a function of the Sn content for different positron mean depths.

**Figure 4.** (a-b) Decomposition of the intensity (a) and lifetime (b) spectra as a function of the positron mean depth for the Ge-VS sample.

**Figure 5.** (a-b) Decomposition of the intensity (a) and lifetime (b) spectra as a function of the positron mean depth for the 13 at.% GeSn sample. (c-d) Decomposition of the intensity (c) and lifetime (d) spectra as a function of the composition at a positron mean depth of 215 nm.

# REFERENCES

[1] L. Pavesi and D.J. Lockwood, editors, *Silicon Photonics III: Systems and Applications* (Springer Berlin Heidelberg, Berlin, Heidelberg, 2016).

[2] R. Soref, Nat. Photonics **4**, 495 (2010).

[3] S. Gupta, B. Magyari-Köpe, Y. Nishi, and K.C. Saraswat, J. Appl. Phys. **113**, (2013).

[4] A. Attiaoui and O. Moutanabbir, J. Appl. Phys. **116**, 063712 (2014).

[5] M.P. Polak, P. Scharoch, and R. Kudrawiec, J. Phys. D. Appl. Phys. **50**, 195103 (2017).

[6] S. Wirths, R. Geiger, N. von den Driesch, G. Mussler, T. Stoica, S. Mantl, Z. Ikonic, M. Luysberg, S. Chiussi, J.M. Hartmann, H. Sigg, J. Faist, D. Buca, and D. Grützmacher, Nat. Photonics **9**, 88 (2015).

[7] J. Aubin, J.M. Hartmann, A. Gassenq, J.L. Rouviere, E. Robin, V. Delaye, D. Cooper, N. Mollard, V. Reboud, and V. Calvo, Semicond. Sci. Technol. **32**, 094006 (2017).

[8] S. Assali, J. Nicolas, S. Mukherjee, A. Dijkstra, and O. Moutanabbir, Appl. Phys. Lett. **112**, 251903 (2018).

[9] W. Dou, M. Benamara, A. Mosleh, J. Margetis, P. Grant, Y. Zhou, S. Al-Kabi, W. Du, J. Tolle, B. Li, M. Mortazavi, and S.-Q. Yu, Sci. Rep. **8**, 5640 (2018).

[10] R.W. Olesinski and G.J. Abbaschian, Bull. Alloy Phase Diagrams **5**, 265 (1984).

[11] R.W. Olesinski and G.J. Abbaschian, Bull. Alloy Phase Diagrams **5**, 273 (1984).

[12] S. Assali, J. Nicolas, and O. Moutanabbir, J. Appl. Phys. **125**, 025304 (2019).

[13] D. Hull and D.J. Bacon, *Introduction to Dislocations* (Elsevier, 2011).

[14] W.P. Bai, N. Lu, A. Ritenour, M.L. Lee, D.A. Antoniadis, and D.-L. Kwong, IEEE Electron Device Lett. **27**, 175 (2006).

[15] M. Christian Petersen, A. Nylandsted Larsen, and A. Mesli, Phys. Rev. B **82**, 075203 (2010).

[16] C.G. Van de Walle and J. Neugebauer, J. Appl. Phys. **95**, 3851 (2004).

[17] É. Bouthillier, S. Assali, J. Nicolas, and O. Moutanabbir, Arxiv.Org/Abs/1901.00436 (2019).

[18] O. Moutanabbir, R. Scholz, U. Gösele, A. Guittoum, M. Jungmann, M. Butterling, R. Krause-Rehberg, W. Anwand, W. Egger, and P. Sperr, Phys. Rev. B **81**, 115205 (2010).

[19] M.O. Liedke, W. Anwand, R. Bali, S. Cornelius, M. Butterling, T.T. Trinh, A. Wagner, S. Salamon, D. Walecki, A. Smekhova, H. Wende, and K. Potzger, J. Appl. Phys. **117**, 163908 (2015).

[20] W. Anwand, G. Brauer, M. Butterling, H.R. Kissener, and A. Wagner, Defect Diffus. Forum **331**, 25 (2012).

[21] A. Wagner, W. Anwand, A.G. Attallah, G. Dornberg, M. Elsayed, D. Enke, A.E.M. Hussein, R. Krause-Rehberg, M.O. Liedke, K. Potzger, and T.T. Trinh, J. Phys. Conf. Ser. **791**, 012004 (2017).

[22] J. Kansy, Nucl. Instruments Methods Phys. Res. Sect. A Accel. Spectrometers, Detect. Assoc. Equip. **374**, 235 (1996).

[23] M. Elsayed, R. Krause-Rehberg, O. Moutanabbir, W. Anwand, S. Richter, and C. Hagendorf, New J. Phys. **13**, 013029 (2011).

[24] N.Y. Arutyunov, M. Elsayed, R. Krause-Rehberg, V. V. Emtsev, G.A. Oganesyan, and V. V. Kozlovski, J. Phys. Condens. Matter **25**, 035801 (2013).

[25] A.F. Makhov, Sov. Phys.-Solid State **2**, (1934).

[26] M. Elsayed, R. Krause-Rehberg, B. Korff, S. Richter, and H.S. Leipner, J. Appl. Phys. **113**, 094902 (2013).

[27] P.J. Schultz and K.G. Lynn, Rev. Mod. Phys. **60**, 701 (1988).

[28] F. Tuomisto and I. Makkonen, Rev. Mod. Phys. **85**, 1583 (2013).

[29] K. Kuitunen, F. Tuomisto, J. Slotte, and I. Capan, Phys. Rev. B **78**, 033202 (2008).

[30] J. Slotte, S. Kilpeläinen, F. Tuomisto, J. Räisänen, and A.N. Larsen, Phys. Rev. B **83**, 235212 (2011).

[31] J. Slotte, K. Kuitunen, S. Kilpeläinen, F. Tuomisto, and I. Capan, Thin Solid Films **518**, 2314 (2010).

[32] M. Elsayed, N. Yu. Arutyunov, R. Krause-Rehberg, V.V. Emtsev, G.A. Oganesyan, and V.V. Kozlovski, Acta Mater. **83**, 473 (2015).

[33] M. Elsayed, N.Y. Arutyunov, R. Krause-Rehberg, G.A. Oganesyan, and V.V. Kozlovski, Acta Mater. **100**, 1 (2015).

[34] R. Krause-Rehberg, M. Brohl, H.S. Leipner, T. Drost, A. Polity, U. Beyer, and H. Alexander, Phys. Rev. B **47**, 13266 (1993).

[35] C. Janke, R. Jones, S. Öberg, and P.R. Briddon, Phys. Rev. B **75**, 195208 (2007).

[36] S. Mukherjee, N. Kodali, D. Isheim, S. Wirths, J.M. Hartmann, D. Buca, D.N. Seidman, and O. Moutanabbir, Phys. Rev. B **95**, 161402 (2017).

[37] I. Riihimäki, A. Virtanen, H. Kettunen, P. Pusa, P. Laitinen, J. Räisänen, and the ISOLDE Collaboration, Appl. Phys. Lett. **90**, 181922 (2007).

[38] S. Decoster, S. Cottenier, U. Wahl, J.G. Correia, and A. Vantomme, Phys. Rev. B **81**, 155204 (2010).

[39] L.I. Khirunenko, M.G. Sosnin, A. V. Duvanskii, N. V. Abrosimov, and H. Riemann, J. Appl. Phys. **123**, 1 (2018).

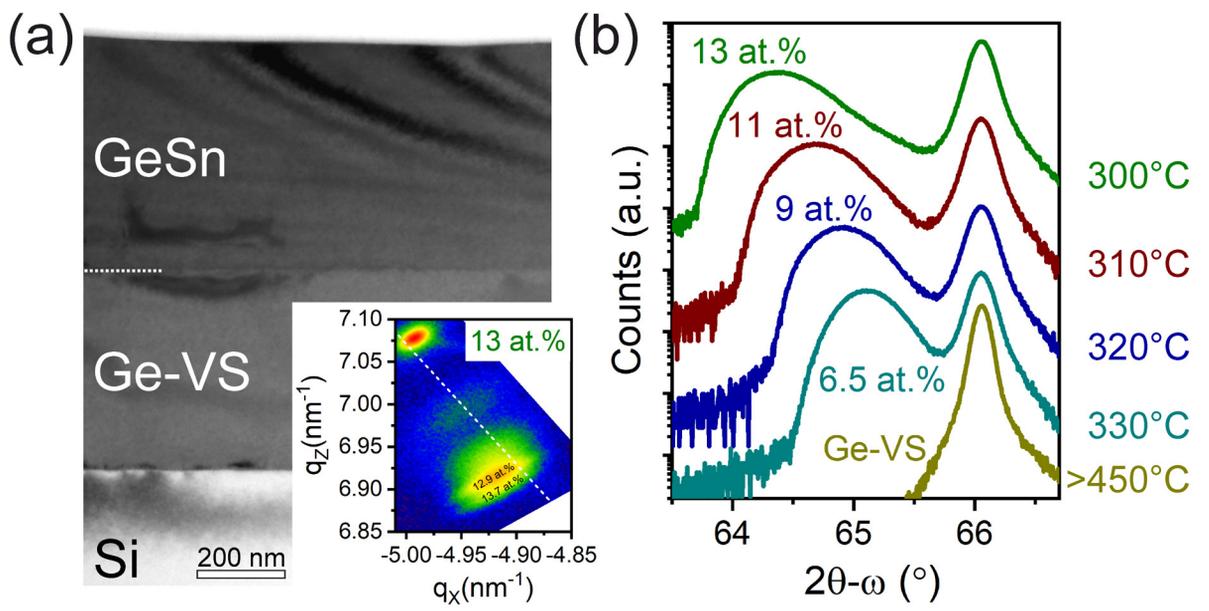

Figure 1

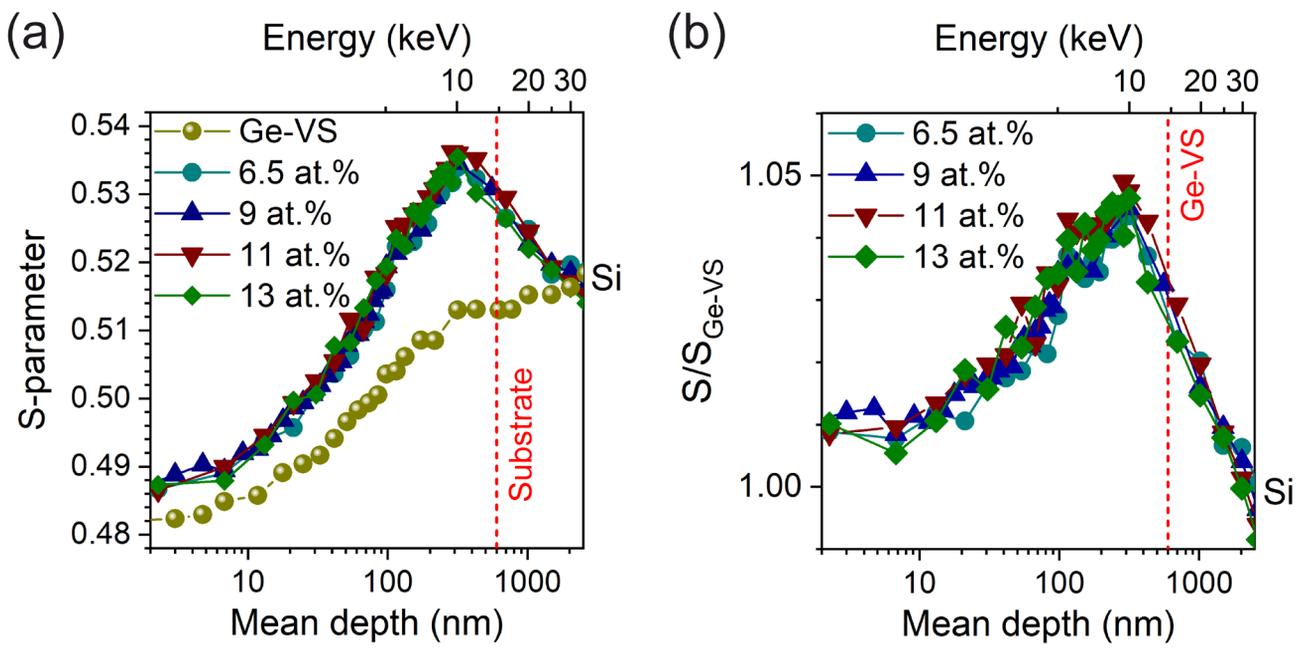

**Figure 2**

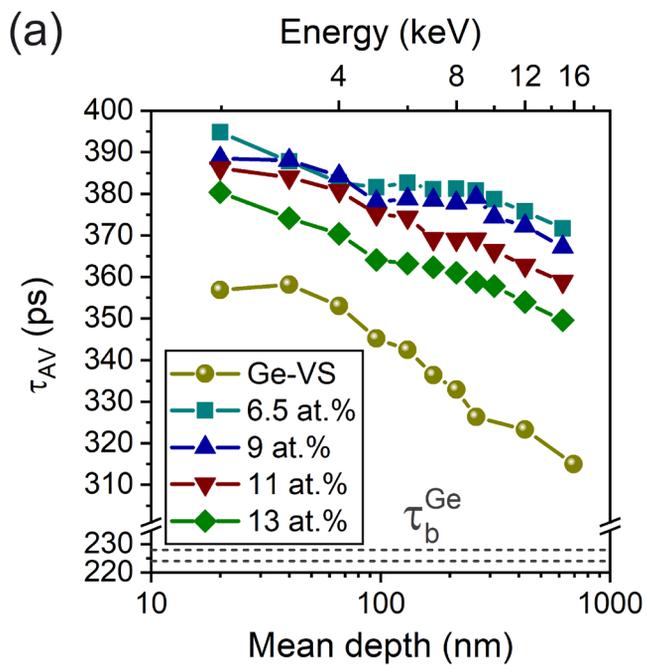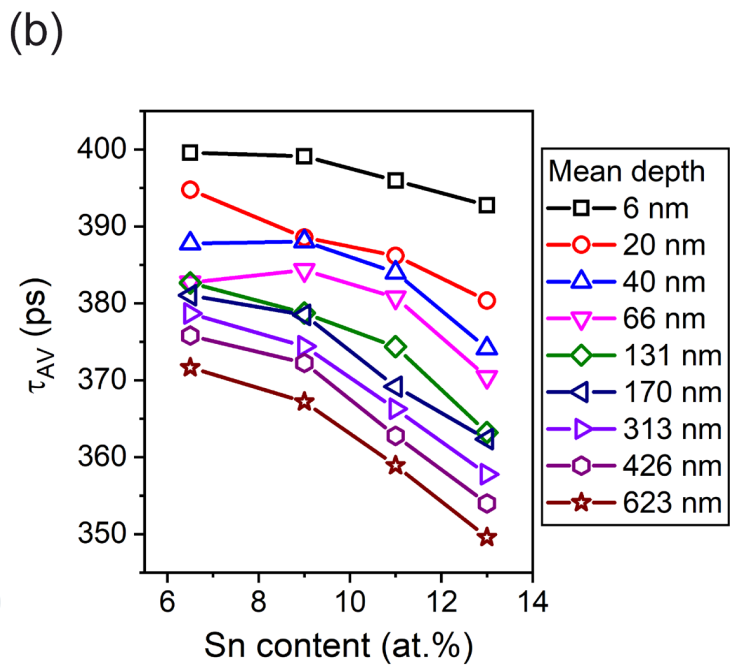

Figure 3

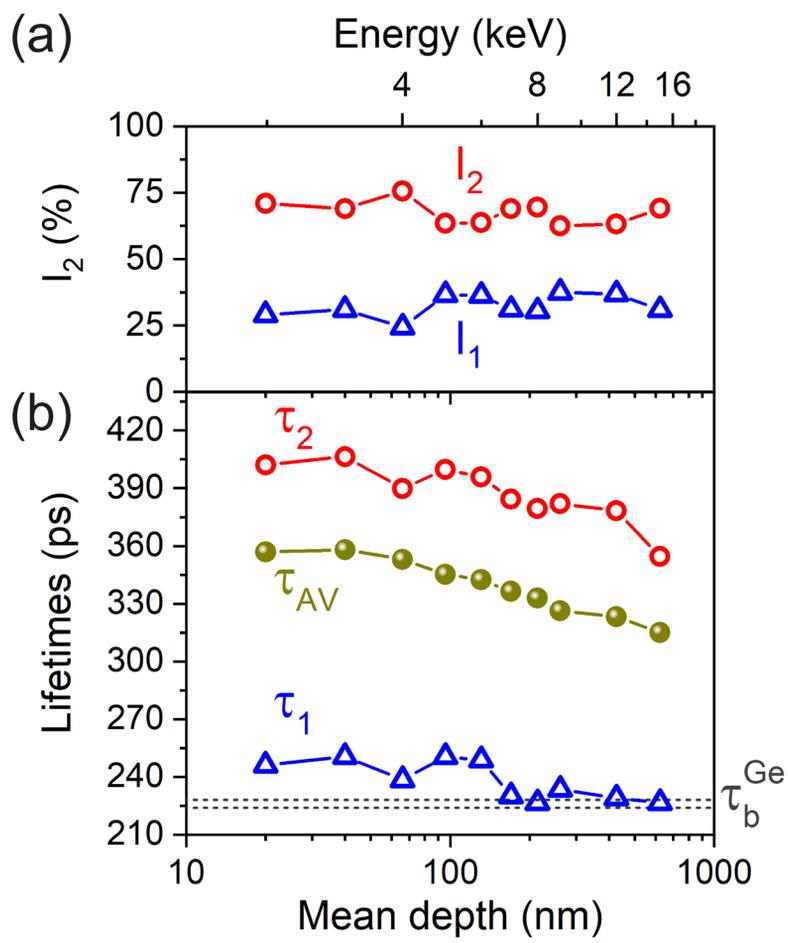

Figure 4

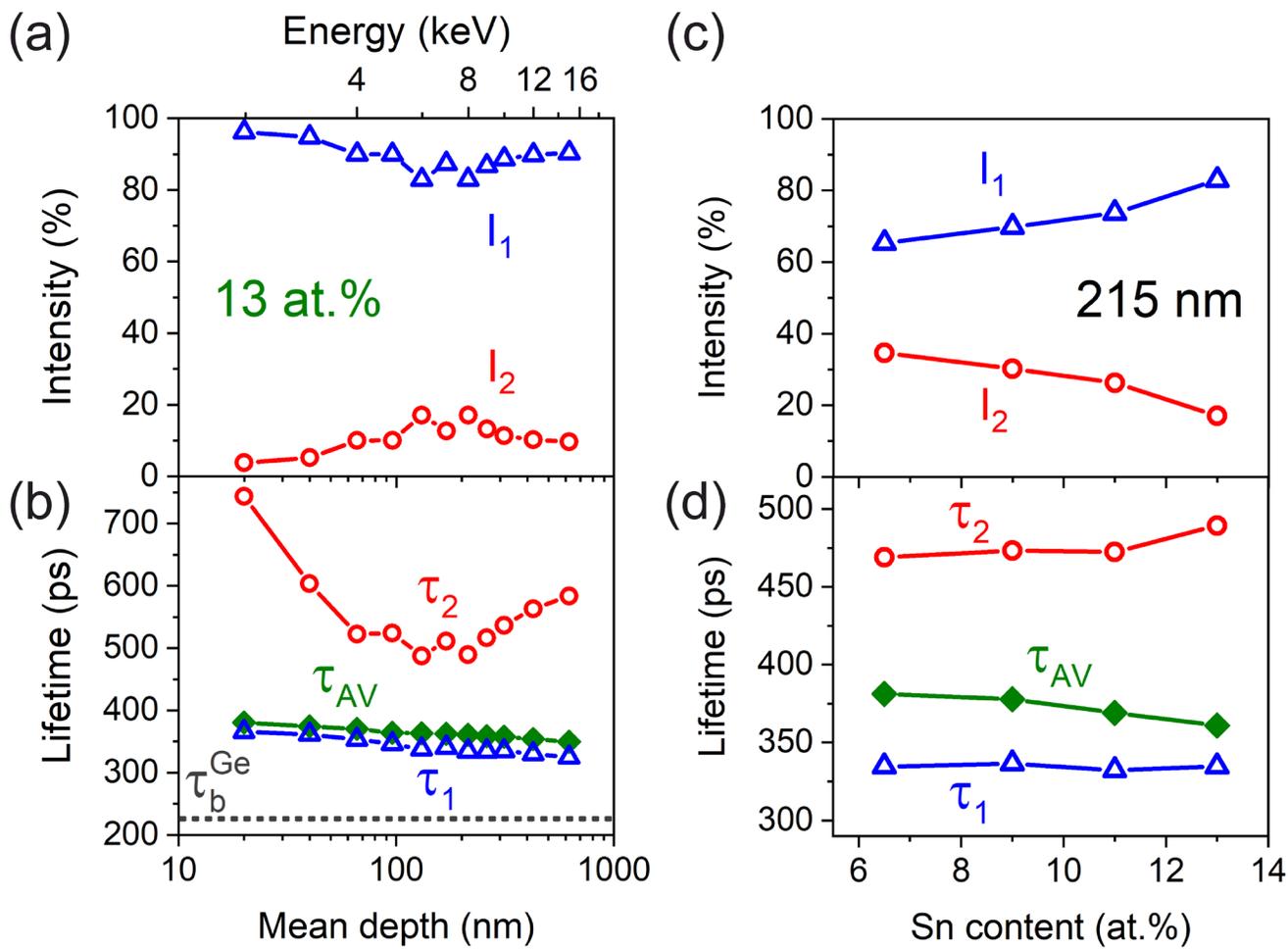

Figure 5